# Characterization of bulk nanobubbles formed by using a porous alumina film with ordered nanopores

*Teng Ma,[†,⊥] Yasuo Kimura,[‡,⊥] Hideaki Yamamoto,[§] Xingyao Feng,[§] Ayumi Hirano-Iwata,[†, §] and Michio Niwano[l,]\**

[†]Advanced Institute for Materials Research (WPI-AIMR), Tohoku University, 2-1-1 Katahira, Aoba-ku, Sendai 980-8577, Japan

[‡]Faculty of Engineering, Tokyo University of Technology, 1404-1 Katakura, Hachioji, Tokyo 192-0914, Japan

[§]Laboratory for Nanoelectronics and Spintronics, Research Institute of Electrical Communication, Tohoku University, 2-2-1 Katahira, Aoba-ku, Sendai 980-8577, Japan

[l]Kansei Fukushi Research Institute, Tohoku Fukushi University, 149-1 Kunimi-ga-oka, Aoba-ku, Sendai 989-3201, Japan

\* niwano@riec.tohoku.ac.jp

[⊥]These authors contributed equally.

Gaseous nanobubbles (NBs), with their unique physicochemical properties and promising applications, have become an important research topic. Generation of monodispersed bulk NBs with specified gas content remains a challenge. We developed a simple method for generating bulk NBs, using porous alumina films with ordered straight nano-scaled holes. Different techniques, such as nanoparticle tracking analysis (NTA), atomic force microscopy (AFM), and infrared absorption spectroscopy (IRAS), are used to confirm NB formation. The NTA data demonstrates that the minimum size of the NBs formed is less than 100 nm, which is comparable to the diameter of nanoholes in the porous alumina film. By generating NBs with different gases, including $CO_2$, $O_2$, $N_2$, Ar, and He, we discovered that the minimum size of NBs negatively correlated with the solubility of encapsulated gases in water. Due to the monodispersed size of NBs generated from the highly ordered porous alumina, we determined that NB size is distributed discretely with a uniform increment factor of $\sqrt{2}$. To explain the observed characteristic size distribution of NBs, we propose a simple model in which two NBs of the same size are assumed to preferentially coalesce. This characteristic bubble size distribution is useful for elucidating the basic characteristics of nanobubbles, such as the long-term stability of NBs. This distribution can also be used to develop new applications of NBs, for example, nano-scaled reaction fields through bubble coalescence.

KEYWORDS: Nanobubble, size distribution, porous alumina film, nanoparticle tracking analysis, infrared spectroscopy, coalescence

## Introduction

NBs are spherical bubbles with a diameter of less than 1 μm. NBs have unique properties such as very long lifetime,[1,2] and high-density surface charge[2-5] and have been actively investigated in many varied applications, including water treatment,[1,6] surface cleaning[7-13] ultrasound contrast agents,[14-17] and biological applications.[14,18] Despite the great potential of NBs in various applications, the basic characteristics of NBs are not fully understood. Various rationales have been proposed for the long-term stability of NBs, [18-20] such as hard hydrogen bonds at the peripheral region (interface) of NBs,[19,21] organic materials adhesion to the surface of NBs,[7] and negative charges generation at the gas/liquid interface of NBs.[5] However, the origin of NB stability still is not completely understood. This stability question remains unanswered partially due to the fact that NBs generated by different methods have different sizes and gas contents, which lead to different properties. Bulk NBs can be generated by electrolysis, though the gas content is limited to oxygen or hydrogen.[22] NBs can also be formed by sudden changes in temperature or pressure of gas-saturated solution, or by pushing gas though a porous glass membrane with random pore diameter.[23-25] However, it is difficult to control the size of the NBs using such methods. To understand the basic characteristics and to explore new applications of NBs, a simple and reliable method for generating monodispersed NBs with any specified gas is needed.

Here, we present a simple technique for generating monodispersed bulk NBs by applying gas pressure to a porous alumina thin film, as schematically shown in Figure 1. A porous alumina film used as a gas filter is formed by anodization. The film has straight, ordered, and densely packed nanoholes with an inner diameter of approximately 100 nm, as is shown in Figure S1(b). Bulk NBs are formed when gas is pushed into the water through the porous alumina nanoholes. These NBs will be monodispersed because the nanohole size of the porous alumina thin film is relatively uniform. Bulk NBs containing various kinds of gas, including rare gases, can be easily produced. To confirm the NB formation by the new method, we used nanoparticle tracking analysis (NTA), infrared absorption spectroscopy in the multiple internal reflection geometry (MIR-IRAS), and atomic force microscope (AFM). We confirmed the generation of gas-filled NBs with a minimum size of less than 100 nm. The minimum NB size depends on the solubility of the encapsulated gases in water. We also found that the size (radius) of NBs takes discrete values, and that those values obey a specific rule. We present a simple model to explain the observed size distribution of NBs. In the model, we assume that bubbles of the same size preferentially coalesce.

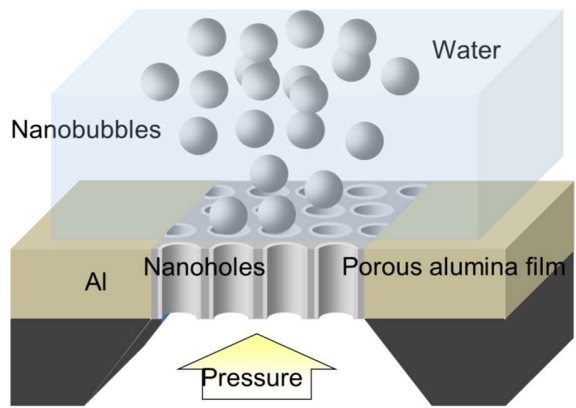

**Figure 1.** Schematic of the proposed method for generating NBs. NBs with almost the same size are formed by applying gas pressure to the porous alumina thin film with straight nanoholes in the film.

**Results and Discussion**

**Confirmation of NB generation.** Porous alumina films used as the NB generator in this study were formed by anodization, as shown in Figure S1. Using the NB generator, we prepared a NB suspension by introducing various gases into a 1-cc solution cell filled with deionized (DI) water (or ethanol) at a gas pressure of 1.5 atmospheres for 1 h. After gas injection, the NB suspension was transferred to a sample cell in the NTA instrument. Under the microscope, the Brownian motion of nanobubbles was clearly visible due to light scattering (Movie S1).

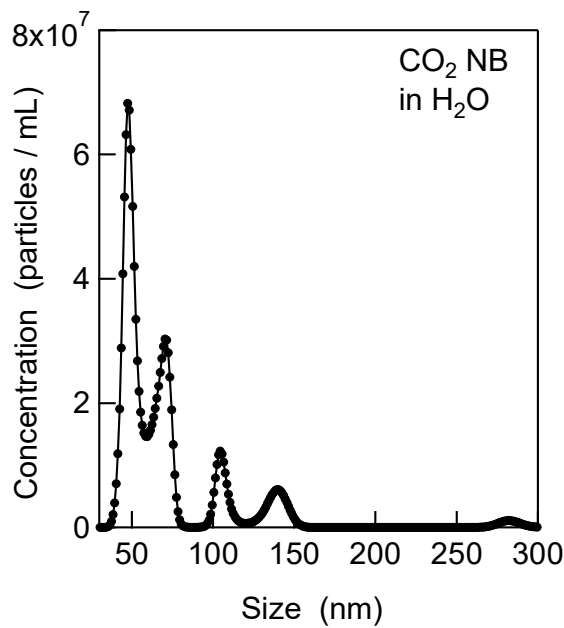

**Figure 2**. Typical size distributions of $CO_2$-containing NBs generated in $H_2O$ measured with NTA method.

In Figure 2, we plot a typical size distribution of NBs prepared by injecting $CO_2$ gas into water. Because NBs move randomly and collide with each other in the suspension, we often observed a size distribution profile with several different sizes, as shown in Figure S2. Distribution profiles in which the NB size is relatively clearly separated, as in the profile in Figure 2, appeared in the ratio of 2 out of 5 sequential measurements, as shown in Figure S2. The discrete size distribution will discuss in detail in the following sections. The NTA data indicates that the size of NBs is in the range of 50 to 200 nm, comparable to the diameter of the straight nanoholes of the porous alumina film used as the gas filter for bubble formation. Thus, we interpret that NBs were generated by passing $CO_2$ gas through the ordered nanohole. As is shown in Figure S3, it was confirmed that $CO_2$ NBs could be generated in ethanol using the proposed method, indicating that NBs can be generated in different kinds of liquid.

To provide additional evidence for the formation of NBs thought the nanoporous alumina film, we carried out MIR-IRAS measurements to examine the $CO_2$ NBs. The experimental setup is shown in Figure S4a. As $CO_2$-NBs were released into DI water, chemical changes in the vicinity of the prism surface were monitored *in-situ*. As already demonstrated by Zhang *et al*., $CO_2$ NBs attached to the prism surface are not stable.[26-28] To stabilize the $CO_2$-filled NBs in water, therefore, we dissolved a surfactant (sodium laureth sulfate) in DI water at a concentration of approximately 150 mg/L before releasing $CO_2$ gas into the water.[29,30] Figure S4b plots a series of IRAS spectra measured while introducing $CO_2$ gas into the surfactant-containing water through the NB generator. We clearly identified absorption bands for gaseous $CO_2$,[29] indicating that $CO_2$-filled NBs were created, and that those NBs gradually gathered in the vicinity of the Si prism surface. After the experiment, we then collected *ex-situ* AFM images of the Si prism surface in contact with surfactant-covered NBs. In Figure S4c, we show typical AFM images of the Si MIR prism surface taken after removing the prism from the solution cell used for MIR-IRAS measurements. The image clearly shows the presence of a number of circular traces on the prism surface. The height of the circular traces is approximately 2 nm, which is comparable to the size of the surfactant used in this study. The diameter of the circular traces is 50 to 100 nm. This diameter is close to that of the NBs observed in the size distribution profiles of Figure 2. Accordingly, we determined that the observed traces originate from the NBs. We conclude that when the NBs on the prism surface broke immediately after the prism was removed from the solution cell, the surfactant that had covered the $CO_2$-containing NBs formed the observed circular traces.

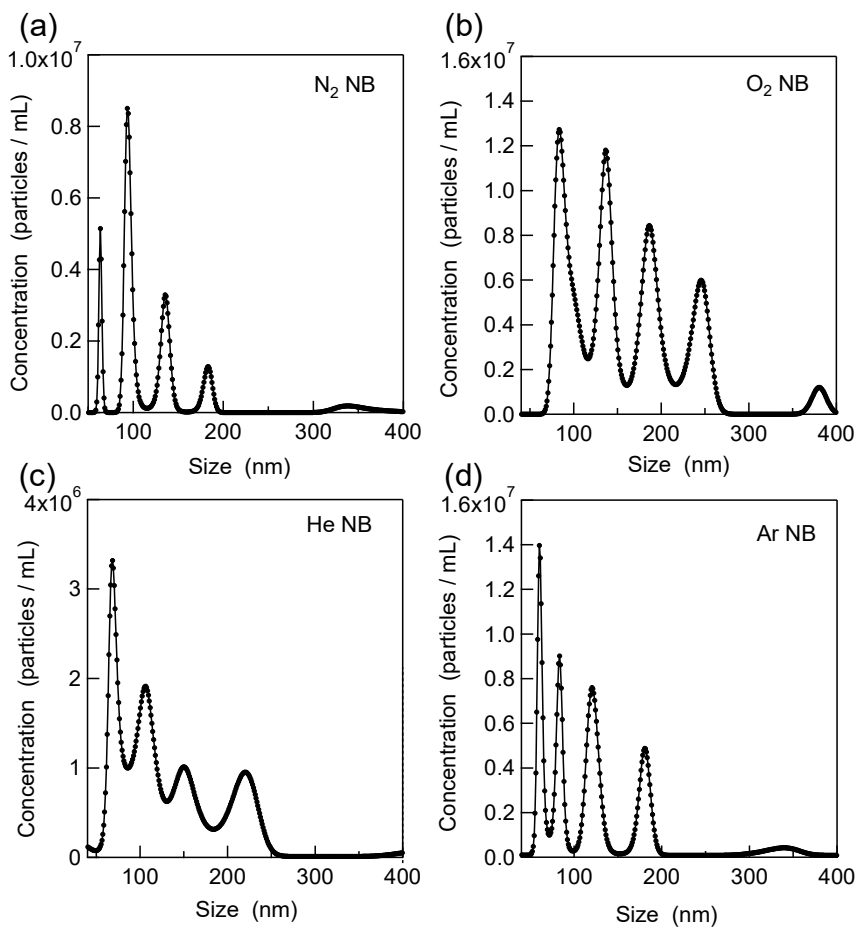

**Figure 3**. The typical size distribution profiles of $N_2$ (a), $O_2$ (b), He (c) and Ar (d) nanobubbles generated in DI water.

**Size distribution of NBs**. In Figure 3, we plot the size distribution of NBs containing different gases. Comparing Figure 3 with Figure 2, we notice that the minimum bubble size of $CO_2$ NB is smaller than NBs of other gases, possibly due to the fact that the solubility of $CO_2$ is higher than that of other gases. To confirm the correlation between the bubble size and the solubility of the encapsulated gas in water, we investigated the relationship between the minimum bubble radius and the solubility of each gas. The results are shown in Figure 4. Gases with higher solubility have smaller minimum NB radius. From this result, we conclude that when NBs emerge from the nanoholes of the porous alumina film, they have a size slightly larger than the diameter of the nanohole, and then shrink, releasing encapsulated gas into the water. The final NB size is determined by the balance between gas solubility and gas permeability of the NB interface. NBs containing $CO_2$ gas, therefore, are smaller than those containing other gases.

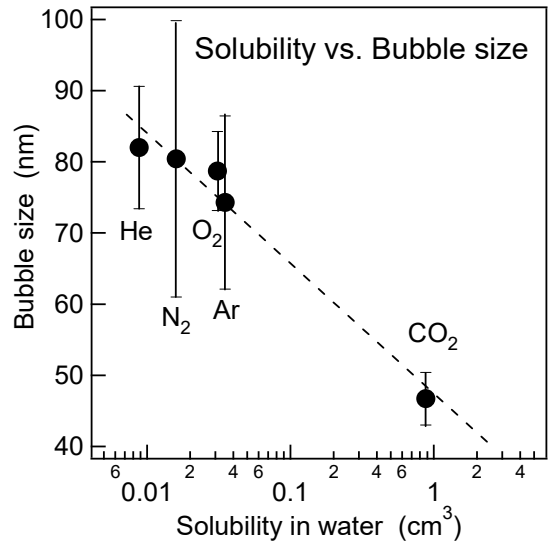

**Figure 4**. Minimum size of NBs as a function of the solubility of the encapsulated gas in water.

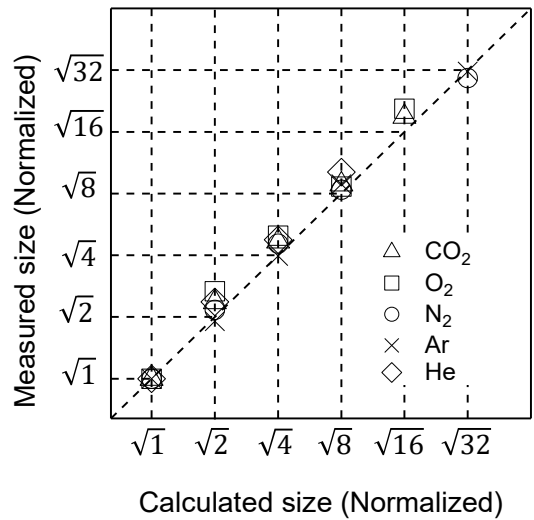

**Figure 5**. Correlation between measured and calculated bubble sizes. The bubble sizes are extracted from the NTA data (peak position) and are normalized by the minimum bubble sizes in every profile. The normalized sizes are all plotted in logarithmic scale.

In Figure 2 and Figure 3, the NB sizes fall into discrete groups. The size distributions of NBs seem to obey a specific rule. For instance, the smallest size of the NBs is approximately 50 nm, and the second smallest size is approximately 75 nm. The ratio of the two sizes is approximately 1.4, which is close to $\sqrt{2}$. The third minimum size is approximately 100 nm that is twice the minimum size. The ratio between adjacent size peaks is approximately $\sqrt{2}$. In order to examine whether such unique size distribution only holds for only $CO_2$ NBs or it is applicable to NBs containing other gases, we plot correlation between measured and calculated bubble sizes (normalized) in logarithmic scale in Figure 5. The result unambiguously shows that the size distributions of NBs obey the specific rule: size ratio between adjacent peaks is $\sqrt{2}$. And this unique size distribution was observed not only for $CO_2$ gas but also for other kinds of gases. This characteristic size distribution was identified regardless of the encapsulated gas of bulk NBs.

**Coalescence of NBs.** As described above, the size of NBs takes discrete values, with an adjacent size ratio of about $\sqrt{2}$. A bubble has a spherical interface between the water and the encapsulated gas. The interfacial energy per unit area is given by the surface tension, $\gamma$. When a bubble shrinks, the interfacial area is reduced, leading to a change in the Gibbs free energy of the system. The change, $\Delta G$, is

$$\Delta G = \gamma \Delta A \qquad (1)$$

where $\Delta A$ is the change in interfacial area.[2] Similarly, it was hypothesized that when two bubbles coalesce the overall interfacial area is reduced, and the energy of the system is reduced. If this is the case, it is likely that, when coalescing, two NBs fuse to produce a single larger NB. To determine the size of the NB formed by coalescence, we consider the case where two NBs with radii of $r_1$ and $r_2$ coalesce to generate a larger NB with a radius of $r_c$. We assume that NB follows the equation of state for an ideal gas, $PV = nRT$, where $P$ is the pressure, $V$ is the volume, $T$ is the absolute temperature, $n$ is the number of moles of gas and $R$ is a universal constant. It is well known that for NBs the pressure, $P$, is represented by the Laplace pressure given by

$$P = 2\gamma/r \qquad (2)$$

where $r$ is the radius of the bubble and $P$ describes the increase in pressure within the bubble with respect to the immediate surroundings. This expression can be derived from a balance between the decrease in surface energy when a bubble shrinks and the increase in internal pressure due to bubble shrinkage: $\sigma \Delta A = p \Delta V$. We furthermore assume that a fixed mass of gas occupies a spherical NB. That is, when the two NBs coalesce, $n_C = n_1 + n_2$, where $n_1$ and $n_2$ are the number of moles of gas of the initial two NBs and $n_C$ is that of the larger NB formed by coalescence. Thus, we can obtain the following relation for the radii of the NBs.

$$r_C^2 = r_1^2 + r_2^2. \qquad (3)$$

This indicates that the overall surface area is conserved upon coalescence of two bubbles, which is contrary to the previously suggested prediction that coalescence will reduce the surface area.[2] When that two NBs before coalescence have the same radius, $r_i \, (= r_1 = r_2)$, we obtain $r_c = \sqrt{2} r_i$. This is in good agreement with our observation that the ratio between adjacent NB sizes is approximately $\sqrt{2}$. Therefore, we suggest that bubbles with the same radius coalesce easily. The NBs created by the porous alumina filter have almost the same radius. As schematically shown in Figure 6, small bubbles of the same size coalesce to form larger bubbles, and the larger bubbles of the same size also coalesce to form even larger bubbles. Such chain coalescence is the origin of the characteristic size distribution profile we observed.

To examine the process of nanobubble coalescence, we look further into our experimental setup. In Figure S1a, the distance between adjacent nanoholes is several tens of nanometers. NBs of the same size are formed because the size of the nanoholes is uniform. Therefore, it is possible that NBs emitted from neighboring nanoholes coalesce almost as soon as they form, resulting in the observed size distribution. We measured the time evolution of NB size distribution. Figure 7 shows the results obtained for the $CO_2$ NB suspension stored in the solution cell of the NTA instrument for different durations. It is evident in Figure 7a-c that, as time passes, NBs of larger sizes are generated, indicating that coalescence gradually proceeded after some time following NB formation. And even after 9 days, the bubble size distribution still obeys the $\sqrt{2}$ increment rule, as is shown in Figure 7d.

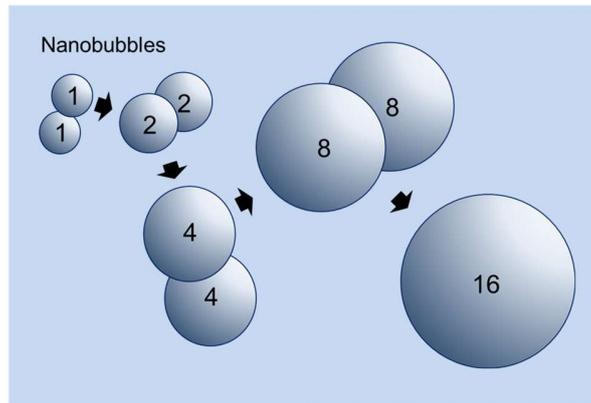

**Figure 6**. Coalescence of NBs. Small bubbles of the same size coalesce to form larger bubbles, and the large bubbles subsequently coalesce to form still larger bubbles. The ratio of the size of a bubble to the smaller bubbles that coalesce to form it is $\sqrt{2}$.

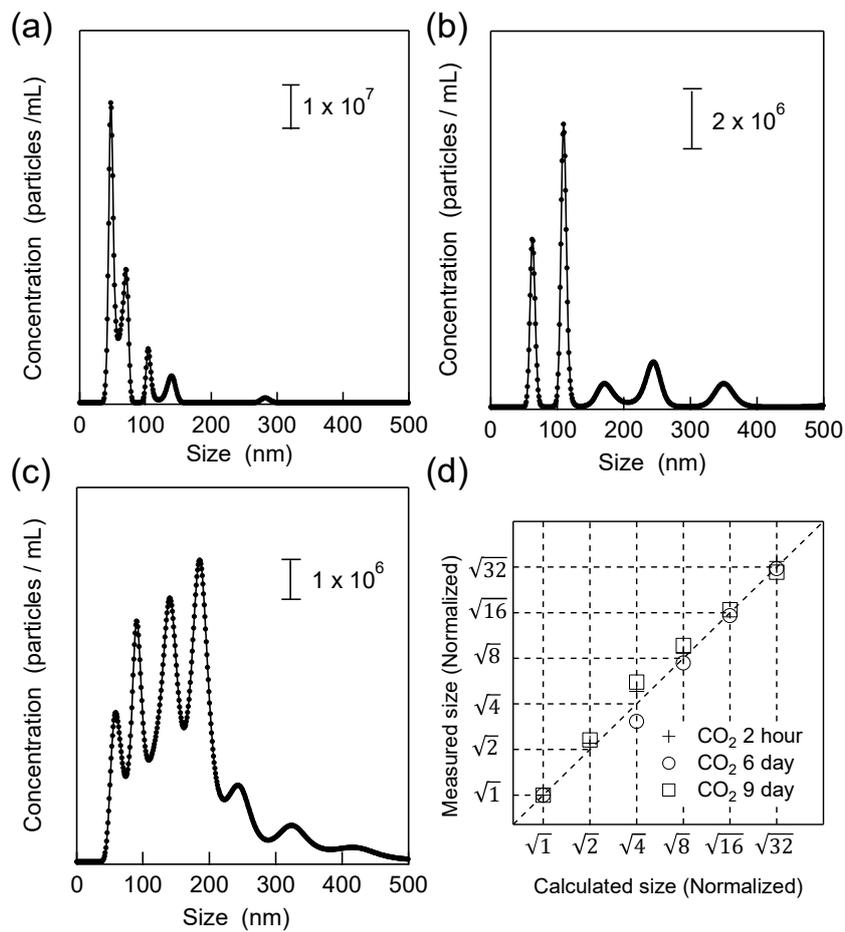

**Figure 7**. Typical size distributions of $CO_2$-containing NBs measured (a) 2 h, (b) 6 days and (c) 9 days after formation of the NB suspension, and the correlation between measured and calculated bubble sizes (d) from 3 profiles.

Why bubbles of the same size are likely to coalesce remains a question. Dahiya et al. reported that the driving force between two micro droplets is at a maximum when the two droplets have the same size.[31] We suspect that the NB may have some similarities with the micro-droplet system. And the driving force between two NBs maximized when they have the same size. Further theoretical and experimental investigations are needed to elucidate the coalescing mechanism.

**Conclusions**

We have developed a simple method to form NBs, using a porous alumina film with straight nanoholes penetrating the film. In order to confirm the NB formation by our method, we used nanoparticle tracking analysis (NTA), infrared absorption spectroscopy in the multiple internal reflection geometry (MIR-IRAS) and atomic force microscope (AFM). NTA data showed that the minimum size of generated NBs is less than 100 nm, which is comparable to the diameter of nanoholes in the porous alumina film. We showed that the minimum size of NBs depends on the solubility of the encapsulated gases in water. We found that the size (diameter) of NBs is discrete, and that there is a specific rule which the bubble size obeys. To explain the observed distribution of the bubble sizes, we proposed a simple model in which two NBs of the same size are assumed to preferentially coalescence. We hypothesized that the repulsive force of two colliding bubbles is smallest when the bubble diameters are the same, which would explain why bubbles of the same size preferentially coalesce.

**Experimental Methods**

**Porous Alumina Film**. Porous alumina films used as the gas filter in this study were formed by anodization. These films have many straight, ordered nano-scaled holes with an inner diameter of approximately 100 nm. Figure S1a shows schematically the fabrication procedure initially described in our previous work.[32] The porous alumina films can withstand an applied pressure of approximately 2 standard atmospheres. A cleaned Al plate (thickness 0.3 mm, purity 99.999%) was electropolished at a constant voltage of 40 V in a mixture of perchloric acid and ethanol (1:4, v/v) at 0 °C for 4 min. Then the Al plate was anodized in 0.3 M oxalic acid for 10 h at 0 °C and an applied voltage of 90 V. The remaining metallic Al on the bottom of the alumina film was removed by etching in 36 wt % hydrochloric acid containing saturated copper(II) sulfate. Finally, the bottom part of the oxide layer (called the barrier layer) was removed by dipping in phosphoric acid at room temperature for 1 h. This final treatment also widened the pore diameter. In Figure S1b, we show typical SEM images (top and bottom views) of the porous alumina film we fabricated. The porous alumina films have nanoholes with a diameter of below 100 nm. Figure S1c is a photograph of the porous alumina film filter used for NB formation in this study.

**Nanoparticle tracking analysis**. Nanoparticle tracking analysis (NTA) has recently been widely used for nanoparticle analysis.[33-35] NTA is a method used to analyze the nanoparticle movement by observing the Brownian motion of each nanoparticle using images taken with an optical microscope. Compared to the widely used dynamic light scattering (DLS) method, NTA allows the measurement of the particle size distribution with high resolution, especially in the range below 1 μm. We measured the size distribution and concentration of NBs in water using a nanoparticle characterization system equipped with a blue polarized 405-nm laser (NanoSight LM10-HSBFT14, Quantum Design Japan). Particles detected by this technique range from 10 to 1000 nm in diameter. The size of NBs, $r$, can be derived using the Stokes-Einstein equation; $D_t = \frac{k_B T}{6\pi\eta r}$, where $k_B$ is the Boltzmann's constant, $T$ is the temperature, is the viscosity of the liquid, and $D_t$ is the diffusion coefficient. The value of $D_t$ is obtained by analyzing the Brownian motion of nanoparticles using NTA. Measurements were carried out at approximately 25 °C in a small cell installed in the NTA equipment. The sampling time is 30 s unless indicated otherwise. Several different locations of the NB-samples were usually measured.

**IRAS measurement.** IRAS detects and distinguishes molecules through infrared absorption spectral profiles of the molecules. Previously, we have shown that biomolecules can be *in-situ* detected in an aqueous solution using MIR-IRAS.[36,37] In the MIR method, a focused infrared light beam passes through a Si prism, internally reflecting many times, and molecular species can be monitored by measuring IRAS spectra of molecules in the vicinity of the prism surface where an evanescent field is present.[38] In Figure S4a schematically depicts the MIR-IRAS experimental setup we used. We attached a Si MIR prism to a small solution cell with a NB generator. NBs produced by the bubble generator were introduced into the solution cell filled with deionized (DI) water. To confirm formation of gas-containing NBs, we monitored *in-situ*, real-time infrared spectral changes of the NB-containing water (NB suspension) in the vicinity of the Si prism surface where an evanescent field of infrared light exists. The Si prism surface was hydrophobized by treating with silane-coupling agents. Bubbles stick easily onto the hydrophobic surface.

ACKNOWLEDGMENT
We are grateful to Dr. Hitoshi Sakamoto and Dr. Ken-ichi Ishibashi for their support in preparation of the porous alumina film filters. We also would like to thank Mr. Hideyuki Saito and Mr. Natsuki Yamada for their support in IRAS measurements. This work was supported by JSPS Grant-in-Aid for Scientific Research (B) (Grant Number: 25600071) and was partly supported by JST CREST (Grant Number: JPMJCR14F3). We would like to thank Editage (*www.editage.com*) for English language editing.

Supporting Information

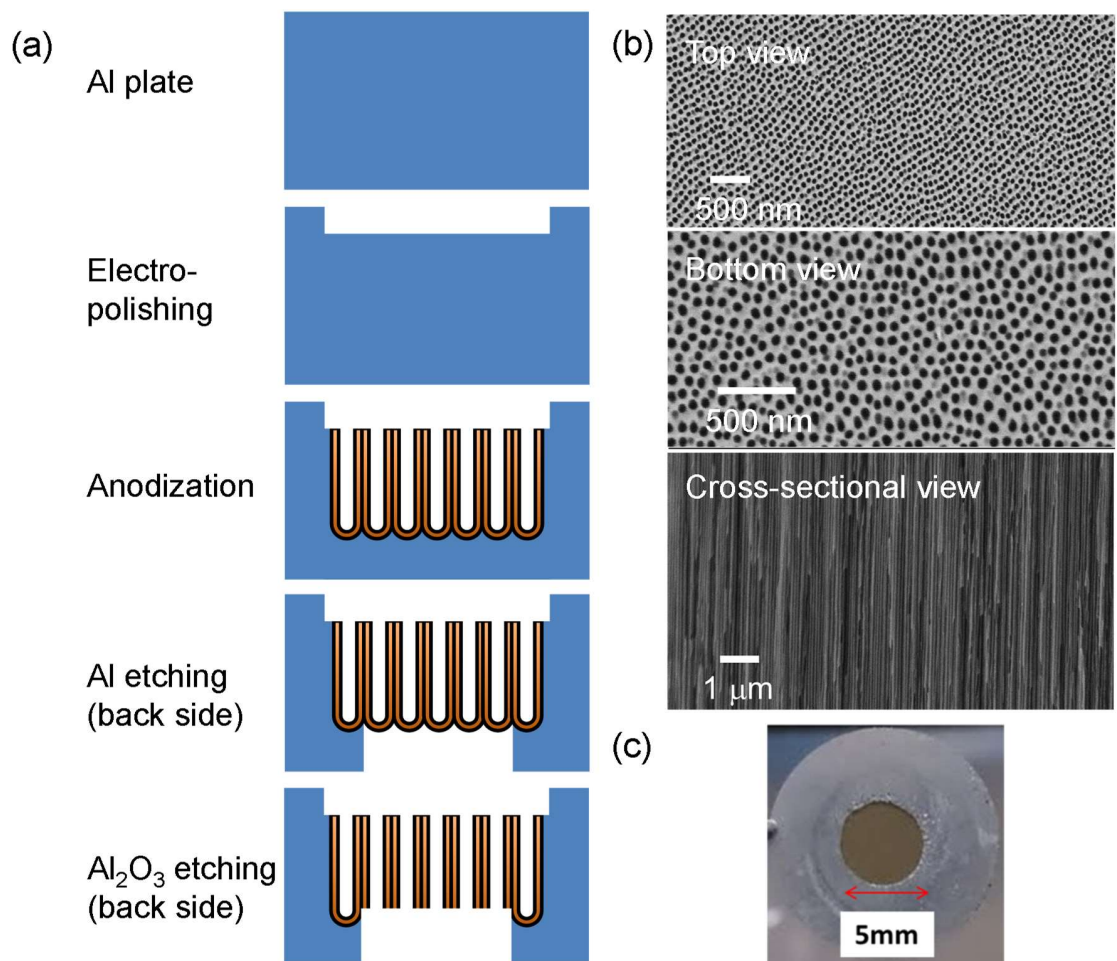

**Figure S1**. (a) Procedure for the fabrication of the porous alumina film filter. (b) Typical SEM images (top, bottom and cross-sectional views) of the porous alumina film. (c) Photograph of the porous alumina film filter used for NB formation (back side).

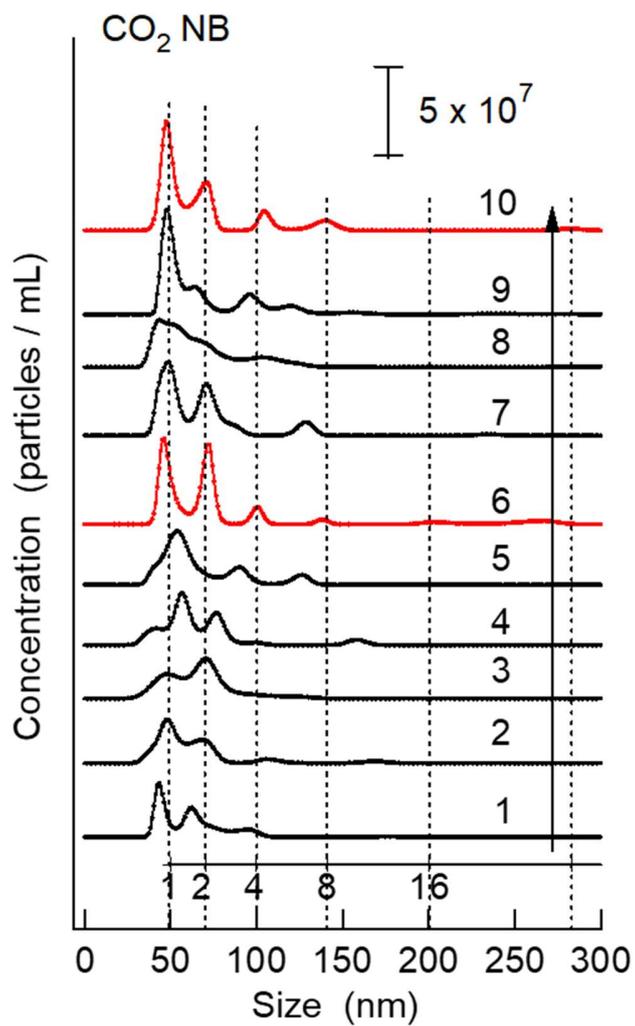

**Figure S2**. Typical size distributions of $CO_2$-containing NBs measured with the NTA method. Because NBs move randomly or contact each other in the suspension, we often observed a size distribution profile with a mixture of different size distributions

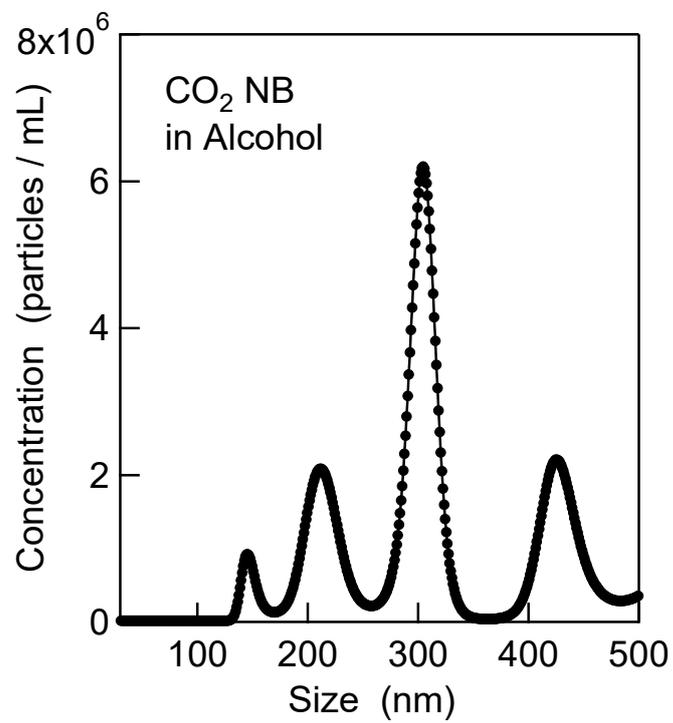

**Figure S3**. Side distribution of $CO_2$-containing NBs generated in ethanol.

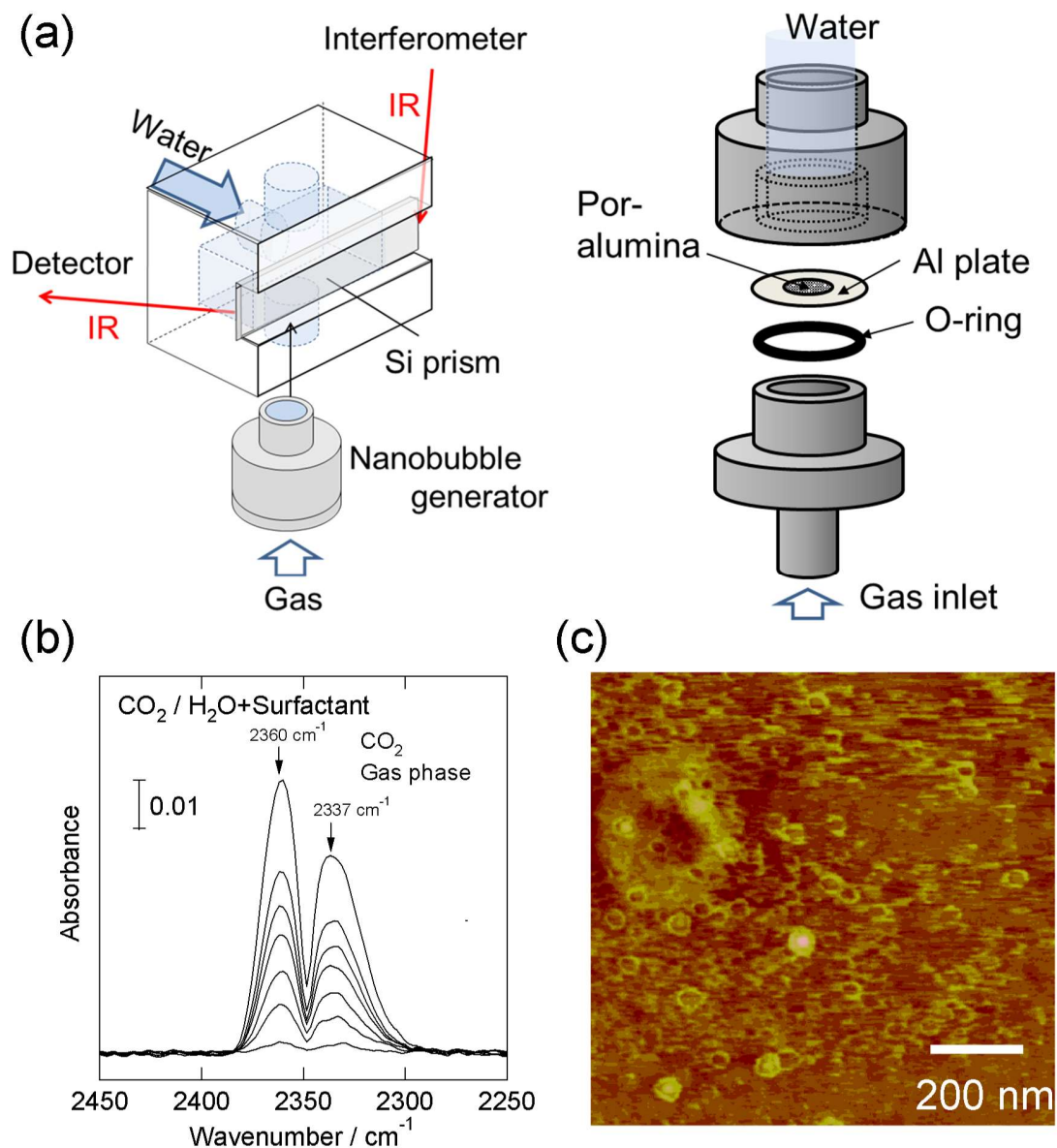

**Figure S4**. (a) Experimental setup for MIR-IRAS measurement and (b) an exploded view of the NB generator with a porous alumina film. (b) *in-situ* measurement of $CO_2$ NBs generated in water with surfactant and (c) *ex-situ* AFM image of the residual of ruptured surfactant-coated-NBs on Si prism